\documentclass[useAMS,usenatbib,fleqn]{mnras}
\usepackage{calrsfs}
\DeclareMathAlphabet{\pazocal}{OMS}{zplm}{m}{n}
\usepackage{multirow}
\usepackage{hhline}
\usepackage{times}
\usepackage[T1]{fontenc}
\usepackage{aecompl} 
\usepackage{amsmath}               
\usepackage{placeins}
\usepackage{booktabs}
\usepackage{amssymb}                
\usepackage{epsfig}                 
\usepackage{epstopdf}
\usepackage{rotating}
\usepackage{color}
\usepackage{tabularx}
\usepackage{url}

\title[Planetesimals into WD discs]{Embedding planetesimals into white dwarf discs from large distances}
\author[Evgeni Grishin \& Dimitri Veras]{
Evgeni Grishin,$^{1}$\thanks{E-mail: eugeneg@campus.technion.ac.il (EG)}
Dimitri Veras,$^{2,3}$\thanks{STFC Ernest Rutherford Fellow}
\\
$^{1}$Technion, Israel Institute of Technology, Haifa, Israel, 3200003\\
$^{2}$Centre for Exoplanets and Habitability, University of Warwick, Coventry CV4 7AL, UK\\
$^3$Department of Physics, University of Warwick, Coventry CV4 7AL, UK
}

\pubyear{2019}

\begin{document}
\label{firstpage}
\pagerange{\pageref{firstpage}--\pageref{lastpage}}
\maketitle

\begin{abstract} 
The discovery of the intact minor planet embedded in the debris disc orbiting SDSS J1228+1040 raises questions about the dynamical history of the system. Further, the recent passage of the potentially interstellar object 1I/'Oumuamua within the solar system has re-ignited interest in minor body flux through exoplanetary systems. Here, we utilize the new analytical formalism from \cite{grietal2018} to estimate the rate at which the gaseous components of typical white dwarf discs trap an exo-planetesimal. We compare the types of captured orbits which arise from planetesimals originating from the interstellar medium, exo-Kuiper belts and exo-Oort clouds. We find that the rate of interstellar medium injection is negligible, whereas capture of both exo-Kuiper and exo-Oort cloud planetesimals is viable, but strongly size-dependent.
For a gaseous disc which extends much beyond its Roche limit, capture is more probable than disruption at the Roche limit. We find that the capture probability linearly increases with the radial extent of the disc.
Even in systems without minor planets, capture of smaller bodies will change the disc size distribution and potentially its temporal variability. Our formalism is general enough to be applied to future discoveries of embedded planetesimals in white dwarf debris discs. 
\end{abstract}




\section{Introduction}

Exoplanetary science is no longer limited to the study of major planets alone. Their smaller cousins -- exo-asteroids and exo-comets -- play vital roles in understanding these systems, particularly after the host star has left the main sequence.

The broken-up remnants of exo-asteroids -- easily seen in the photospheres of white dwarfs (WDs) -- provide us with the most direct and abundant measurements of exoplanetary building blocks \citep{gaeetal2012,juryou2014,haretal2018,holetal2018}. These minor planets also reveal insights about planetary architectures, having been gravitationally perturbed towards the WD through a variety of dynamical mechanisms \citep{veras2016}.  These mechanisms act in single-star systems with a single planet \citep{bonetal2011,debetal2012,frehan2014} or multiple planets \citep{veretal2016a,musetal2018,smaetal2018}, which may or may not harbour moons \citep{payetal2017}. The presence of a stellar binary companion alone can trigger such perturbations, even without the presence of a planet \citep{hampor2016,petmun2017,steetal2017}.

All types of planets which reach the WD phase must have survived engulfment from giant branch evolution. Two critical engulfment distances exist: one for the red giant branch phase \citep[e.g.][]{viletal2014,galetal2017,raoetal2018,sunetal2018}
and another for the asymptotic giant branch phase \citep{musvil2012}. Some treatments are general enough to determine the critical distances for both phases \citep{adablo2013,norspi2013,madetal2016}.

Before accreting onto a typical $R_{\oplus}$-sized WD, these minor planets will enter the star's Roche, or disruption, radius, which extends out to about $1R_{\odot}$. Although direct infall onto the WD is possible, that target is small relative to its Roche radius. Hence, the much more likely outcome is the production of a debris disc \citep{graetal1990,jura2003,debetal2012,beasok2013,veretal2014,veretal2015}. The size distribution which feeds the disc is unknown but may be constrained \citep{wyaetal2014,broetal2017,kenbro2017a,kenbro2017b}. The compactness and other physical properties of these discs have been investigated in only a couple handfuls of investigations \citep{bocraf2011,rafikov2011a,rafikov2011b,metetal2012,rafgar2012,mirraf2018}.

The theory of WD debris disc formation is bourne out by observations. Of the over 40 discs known \citep{farihi2016}, about 20 per cent contain observable gas as well as dust \citep[e.g.][]{ganetal2006,denetal2018}. We highlight three of these discs which have multi-epoch observations here: those orbiting SDSS J1228+1040, HE 1349-2305 and WD 1145+017. The structure of the first two can be deduced from their emission line profiles \citep{ganetal2006,manetal2016,manetal2019}, whereas the third is seen in absorption \citep{redetal2017,cauetal2018}. These discs, with orbital periods of $\approx$ 5 hours, have been observed for about 1-10 years, and are suggestive of dynamically active systems.

In fact, individual exo-asteroids and smaller planetesimals or fragments have been observed in two of them. A single exo-asteroid has been observed to be distinegrating in real time for the last four years around WD 1145+017 \citep{vanetal2015,vanrap2018}, shedding debris that contributes to the structure and evolution of the accompanying disc. A minor planet of a very different type that is consistent with a planetary core fragment has been detected orbiting SDSS J1288+1040 \citep{manetal2019}: this minor planet is embedded within the disc, at about $0.73 R_{\odot}$ -- which is well within the canonical $1 R_{\odot}$ value -- and does not feature observable ablation. Although no planetesimal has so far been discovered orbiting HE 1349-2305 \citep{denetal2018}, deeper observations are warranted.

These known exo-minor planets range in size from km-scale to hundreds of km \citep{rapetal2016,veretal2016b,garetal2017,guretal2017,veretal2017,manetal2019}. However, many smaller fragments of unconstrained radii appear and disappear regularly in the photometric transit curves of the debris around WD 1145+017 \citep{vanetal2015,ganetal2016,rapetal2016,izqetal2018}. Overall, planetesimals ranging in size of many orders of magnitude are either embedded within or just outside of WD discs.

However, despite their importance in exoplanetary science, these minor bodies still have murky dynamical histories. Even more challenging is the explanation of some of their current near-circular \citep{veretal2017} or slightly or moderately eccentric orbits \citep{manetal2019}. In this paper, and inspired by the minor planet orbiting SDSS J1228+1040 and the recent passage of the interstellar interloper 1I'/'Oumuamua through the solar system \citep{meeetal2017}, we consider the probability of capturing interstellar and exo-planetesimals due to gas drag with a scaled-down gaseous disc.

Recently, \cite{grietal2018} showed that passing interstellar medium (ISM) planetesimals could be captured via aerodynamic gas drag in a protoplanetary disc. The number of 'Oumuamua-like planetesimals which are captured in a protoplanetary disc over its lifetime could is around $10^3-10^5$, depending on the stellar environment with respect to the kinematics of field stars and clusters. In addition, for typical planetesimal size distributions, a few of the captured objects could be $\gtrsim$ km-sized. 

Although the total mass captured is small compared to the dust abundance, the seeding of large planetesimals could catalyze planet formation in those discs and alleviate many of the problems involved in the formation of the first planetesimals, while overcoming the metre-size barrier. Perhaps similarly, there are also potential ``second-generation'' formation implications for objects embedded inside WD discs \citep{schdre2014,voletal2014,hogetal2018,vanetal2018}.

In this paper, we estimate the rate and number of captured ISM planetesimals, exo-Kuiper belt and exo-Oort cloud objects in a WD planetary remnant debris disc. In Section \ref{s2}, we describe our assumptions of the WD gaseous disc properties. We derive the capture probabilities, rates and total number of captures in Section \ref{sec:ISMprob}. Section \ref{lc} features loss cone dynamics for the exo-Oort cloud case. We discuss the implications in Section \ref{sec5} and summarize in Section \ref{sec6}.

\section{Discs around White Dwarfs} \label{s2}




The lifetimes of WD debris discs are unknown. However, those can be estimated by dividing the the average mass of metals within the convection zone of DB WDs by the averaged accretion rates found in DA WDs. The result is $3 \times 10^4 - 5 \times 10^6$ yr \citep{giretal2012}. Given the lifetimes of these discs and the tidal dissipation timescales, the amount of mass accreted by an embedded planetesimal is unknown.

The eight known WD debris discs which contain both gas and dust are kinematically constrained to have inner radii of about $0.6 R_{\odot}$ and outer radii of about $1.2 R_{\odot}$. Their masses and size distributions are unknown, but between the inner rim and the WD photosphere any material transferred from the disc to the WD must be sublimated in the form of gas.

The structure of the gaseous components of WD discs has been explored by \cite{metetal2012} and \cite{kenbro2017b}. Both sets of authors assume a constant arrival rate of solids onto the Roche radius and subsequent sublimation, but different temperature profile and $\alpha$-viscosity prescriptions.  The solid component drifts inward due to Poynting-Robertson (PR) drag, while the gaseous component is spread out by viscous evolution. The inner region accreting directly onto the WD, while the outer region extends far beyond the Roche radius, up to $\approx 20\  R_{\odot}$. In fact, the outward spreading is constrained only by the location at which solids recondense.

Both models predict a co-existent state of gas and debris, which is supported by observations \citep[e.g.][]{ganetal2006,farihi2016,denetal2018}. It remains unclear why the gas does not condense back into solids beyond the sublimation point. \cite{vanetal2018} assumed that once the gas exceeds the Roche radius, it condenses to form minor planets. It is unclear why the unknown process that prevents condensation should be efficient only inside the Roche radius, namely where for optically thick discs the temperature is independent of the radius, as in \cite{metetal2012}. Here we assume a constant feeding rate, similarly to \cite{metetal2012,kenbro2017b}, and that condensation is stalled also on larger scales and extend the gaseous disc much further away than the Roche radius. 
\subsection{Gaseous disc structure} \label{sec:2.1}


We assume for simplicity that the WD gaseous discs
extend from $r_{{\rm in}}
=8.7\cdot10^{8}$ cm to $r_{{\rm out}}=1.4 \cdot 10^{12}$ cm $\approx 20 R_{\odot}$,
spanning three orders of magnitude. We normalize the surface density
at $r_{0}\equiv10^{10}$ cm, such that the surface density profile
is 
\begin{equation}
\Sigma_{g}(r)=\Sigma_{g,0}\left(\frac{r}{r_{0}}\right)^{-\beta}
\end{equation}
where $\beta$ is an arbitrary exponent. \cite{metetal2012} find that $\Sigma_{g}(r) \propto r^{-n-1/2}$, where $n$ describes the viscosity power law, $\nu (r) \propto r^n$. For a flat temperature profile, $n=3/2$, and $n=1$ for optically thin disc with $T(r) \propto r^{1/2} $. We thus expect that $\beta$ will be in the range of $1.5-2.0$.  

The total mass of the disc ($\beta \ne 2$)
is 
\begin{align}
M_{{\rm disc}} & =\intop_{r_{{\rm in}}}^{r_{{\rm out}}}\Sigma_{g}(r)2\pi rdr=2\pi\Sigma_{g,0}r_{0}^{\beta}\intop_{r_{{\rm in}}}^{r_{{\rm out}}}r^{1-\beta}dr\\ \nonumber
 & =\frac{2\pi}{2-\beta}\Sigma_{g,0}r_{0}^{2}\left[\left(\frac{r_{{\rm out}}}{r_{0}}\right)^{2-\beta}-\left(\frac{r_{{\rm in}}}{r_{0}}\right)^{2-\beta}\right]
\label{Mdisc1}
\end{align}
For $\beta<(>)2$ most of the mass is in the outer (inner) regions. In this case the total mass is determined by the outer (inner) boundary
\begin{equation}
M_{{\rm disc}}\approx\frac{2\pi}{|2-\beta|}\Sigma_{g,0}r_{0}^{2}\left(\frac{r_{{\rm io}}}{r_{0}}\right)^{2-\beta},
\end{equation}
where $r_{\rm io}$ is the outer (inner) radius id $\beta <(>)2$.

For $\beta=2$, each radial octave contributes the same amount of mass, and the total disc mass depends on the Coulomb 
logarithm $\ln (r_{\rm in}/r_{\rm out})$ as
\begin{equation}
M_{\rm disc}=2\pi\Sigma_{g,0}r_{0}^{
2}\ln
\left(
\frac{r_{\rm out}}{r_{\rm in}}
\right).
\label{Mdisc3}
\end{equation}

By inserting the fiducial values and taking $M_{{\rm disc}}=10^{25}$g
with representative values of $\beta$, we have 
\begin{equation}
\Sigma_{g,0}=10^{3}\rm{g\ cm^{-2}}\left(\frac{r_{0}}{10^{10}{\rm cm}}\right)^{-\beta}\left(\frac{M_{{\rm disc}}}{10^{25}{\rm g}}\right)\begin{cases}
0.68 & \beta=3/2\\
2.35 & \beta=5/2\\
51 & \beta=2 \label{eq:sigma0}
\end{cases}
.
\end{equation}

To summarize, the typical normalization for a $10^{25}$g gas disc
is $\sim10^{3} - 10^{4}\ {\rm g\ cm^{-2}},$ regardless of the power-law density
exponent $\beta$.


\section{Capture Rates} \label{sec:ISMprob}

The capture rates of different populations of planetesimals depend both on the supply rate, and on the capture probability. For the ISM planetesimals, the supply rate is constant and depends mainly on the environment, while the capture probability is highly dependent on the size of the planetesimals and the disc structure. Conversely, exo-asteroid and exo-Oort objects are captured efficiently once they pass close enough to the disc, as long as they are not too massive. The supply rate, on the other hand, depends on the diffusion rate that brings new planetesimals into orbits that are lost, leading to the loss-cone dynamics presented in detail in sec. \ref{lc}.

\subsection{ISM planetesimals}

Consider a planetesimal of size $R_p$ and density $\rho_p$, on a hyperbolic orbit with velocity at infinity $v_{\infty}$ and impact parameter $b$, approaching a WD with mass $M_{\rm WD}$ and harbouring the gaseous disc structure that was described in sec. \ref{s2}. A planetesimal passing through the disc loses energy due to aerodynamic gas drag. The most massive planetesimals could be decelerated by dynamical friction at lower relative velocities \citep{g15, g16}, but the initial capture occurs at high relative velocities, where dynamical friction in inefficient.

The planetesimal remains bound if the total loss energy is greater than its kinetic energy at infinity. \cite{grietal2018} found that this capture condition is 
\begin{equation}
R_{p}\lesssim\frac{3}{4}\frac{C_{D}\Sigma_{g}(q)}{\rho_{p}}\left(1+\Theta_{s}\right),\label{eq:capture_cond}
\end{equation}
where $C_D$ is the drag coefficient, $\Theta_{s}\equiv v_{\rm esc}^{2}/v_{\infty}^{2}$ is the gravitational
focusing Safronov number and $v_{\rm esc}^{2}=2GM_{\rm WD}/q\approx v_{\rm rel}^{2}-v_{\infty}^{2}$
is the escape velocity at closest approach $q$. For $\Theta_{s}\gg1$
gravitational focusing is important, while for $\Theta_{s}\ll1$ the
scattering is mostly in the geometric collision regime. 
 
By assuming that the velocities are distributed from a Maxwellian velocity distribution with dispersion $\sigma$, and the impact parameters are distributed from an area-uniform impact parameter distribution with cut-off $b_{\rm max}$, \cite{grietal2018} found analytic expressions for the capture fractions as a function of the planetesimal's size and disc parameters. Although small pebbles are dominated by the geometric regime,  large planetesimals are dominated by gravitational focusing. For the geometric regime, the capture probability is 
\begin{equation}
f^{\rm geo}_{c}(R_{p})\approx\left(\frac{3C_{D}\Sigma_{g,0}}{4\rho_{p}R_{p}}\right)^{2/\beta}\left(\frac{b_{\rm max}}{r_0}\right)^{-2},\label{eq:fc1}
\end{equation}

\noindent{}while for the gravitational focusing regime, the capture probability is

\begin{equation}
f^{\rm gf}_c(R_p) =   \Gamma(\beta')\left(\frac{3C_{D}\Sigma_{g,0}}{4\rho_{p}R_{p}}\right)^{\frac{1}{1+\beta}} \left(\frac{GM_{\rm WD}}{\sigma^{2}{r_0}}\right)^{\frac{2+\beta}{1+\beta}} \left(\frac{{r_0}}{b_{{\rm max}}}\right)^{2}, \label{eq:fc2}
\end{equation}
where $\beta' = \beta / (1+\beta) $, and $\Gamma(x)$ is the Gamma function.

\begin{figure}
	\includegraphics[width=\columnwidth]{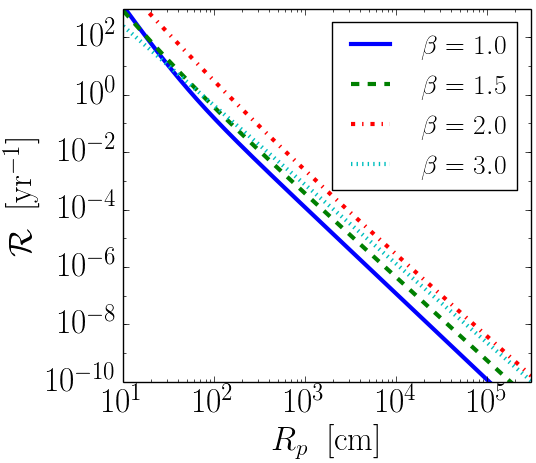}
    \caption{Total rates of interstellar objects embedded in the disc as a function of object size. Each line plots Eq. (\ref{eq:rate}) with $\alpha=11/6$, $f=1$, $M_{\rm WD} = 0.6M_{\odot}$, and $M_{\rm disc} = 10^{25} \ \rm g$. The disc profile exponent is $\beta = 1, 1.5, 2$ and $3$ for the solid blue, dashed green, dot-dashed red and dotted cyan lines, respectively.
    }
    \label{fig:rates1}
\end{figure}

Multiplying the capture probabilities (Eq. \ref{eq:fc1} and \ref{eq:fc2}) by $n_{\rm ISM}(\sqrt{8/\pi}\sigma) \pi b_{\rm max}^2$ gives the total capture rate.  Here, $n_{\rm ISM}$ is the number density of ISM planetesimals of size less than $R_p$, which depends on the mass function and overall density. Here, we assume there is at least $fM_{\rm \oplus}$ mass in ejected planetesimals, where $f=1$.

The number density of planetesimals is $n_{\rm ISM} = n_{\star} N(R_p)$, where $n_{\star} = 0.1 \ \rm pc^{-3}$ is the stellar number density in the field and $N(m)$ is the number of planetesimals of mass~$\le~m$, which are estimated as follows:  The mass function is likely to follow a power law distribution, $dN/dm \propto m^{-\alpha}$. Small bodies are likely to follow collisional (Dohnanyi) distribution, \citep{d69} i.e. $dN/dm \propto m^{-11/6}$, while large bodies could by formed by streaming instability, i.e. $dN/dm \propto m^{-5/3}$. We leave the mass function exponent $\alpha$ as a free parameter. The number of planetesimals is \citep{grietal2018}

\begin{equation}
N(m)=\frac{2-\alpha}{\alpha-1}10^{2-\alpha}f\left(\frac{R_p}{R_\oplus} \right)^{3-3\alpha}, \label{eq:nm}
\end{equation}
where $R_{\oplus}$ is Earth's radius. In Eq. (\ref{eq:nm}) we assumed that the largest ejected planetesimal mass is $0.1M_{\oplus}$.

The total capture rates are thus 
\begin{equation}
\mathcal{R}(R_p) = \frac{2-\alpha}{\alpha-1}10^{2-\alpha}\sqrt{8\pi}f\left(\frac{R_p}{R_\oplus} \right)^{3-3\alpha} K(R_p) n_{\rm \star}\sigma r_0^{2}, \label{eq:rate}
\end{equation}
where
\begin{equation}
K(R_P) \equiv \begin{cases} \left(\frac{3C_{D}\Sigma_{g,0}}{4\rho_{p}R_{p}}\right)^{2/\beta} & \rm geo \\
\Gamma\left(\frac{\beta}{1+\beta}\right)\left(\frac{3C_{D}\Sigma_{g,0}}{4\rho_{p}R_{p}}\right)^{1/(1+\beta)} \left(\frac{GM_{\rm WD}}{\sigma^{2}{r_0}}\right)^{\frac{2+\beta}{1+\beta}} & \rm gf
\label{eq:kkk}
\end{cases}
\end{equation}

Figure \ref{fig:rates1} shows the total disc-embedding rates obtained by Eq. (\ref{eq:rate}). The typical values are of a system of a WD of mass $M_{\rm WD} = 0.6M_{\odot}$, and a total mass of the disc $M_{\rm disc} = 10^{25}\ \rm g$ with surface density profile exponent $\beta$ as a free parameter. The normalized surface density $\Sigma_{g,0}$ is set from Eq. (\ref{eq:sigma0}) in sec. \ref{sec:2.1}. The solid density is set to $\rho_p = 1\ \rm g\  cm^{-3}$. 

The total mass capture rate scales as $\sim \mathcal{R} \rho_p R_p^3 $, and ranges around $10^{-4} - 10^{-2}\ \rm g\  s^{-1} $, around 10 orders of magnitude less than the inferred metal pollution rate of $\sim 10 ^{8} \ \rm g s^{-1}$. Thus, ISM capture cannot be the source of metal pollution on WDs, as expected. Nevertheless, the captured ISM material could play a role in the overall disc evolution, similarly to protoplanetary discs. In addition, the frequent capture of cm-sized pebbles may provide observational signatures.

\subsection{Exo-Kuiper belt objects and exo-Oort cloud comets}
\begin{figure*}
    \begin{centering}
	\includegraphics[width=\columnwidth]{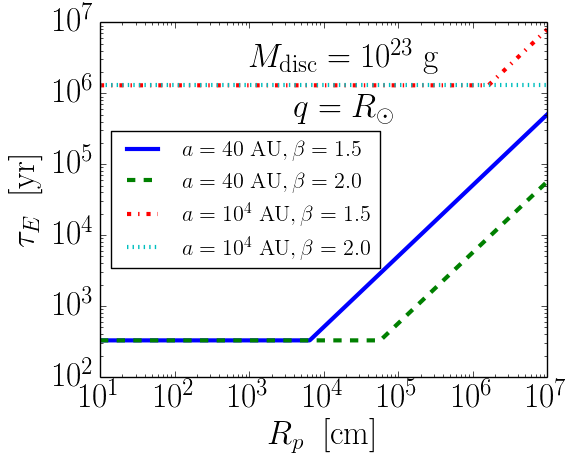}
    \includegraphics[width=\columnwidth]{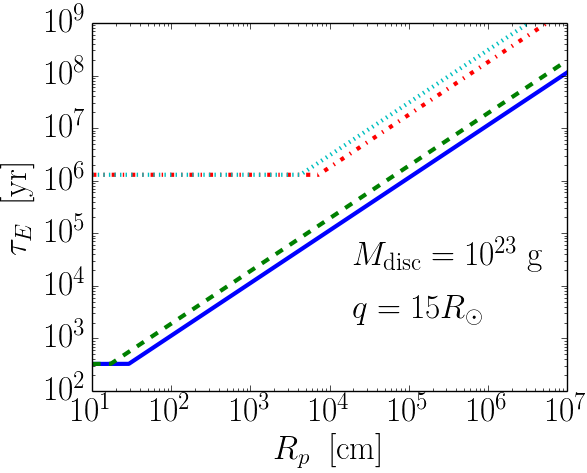}
    \includegraphics[width=8.5cm]{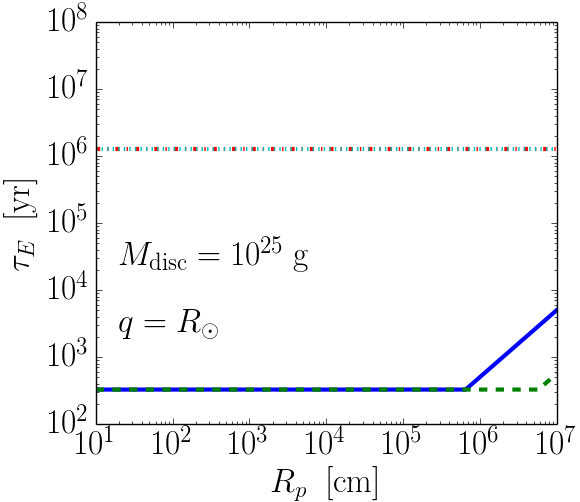}
    \includegraphics[width=8.1cm]{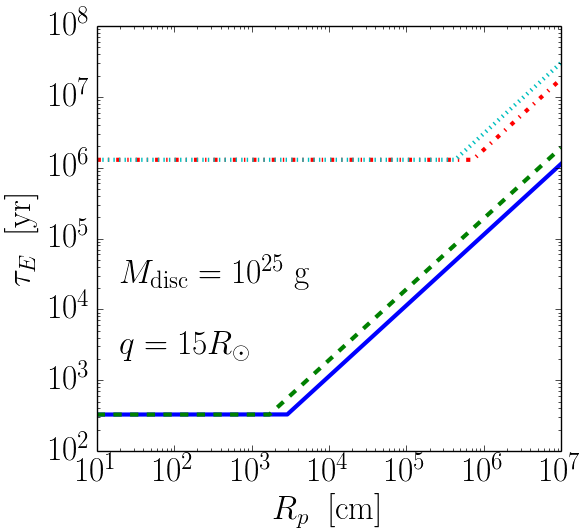}
    \par\end{centering}
    \caption{Dissipation times (times to achieve final orbit) of bound exo-Kuiper Belt objects and exo-Oort cloud objects. Each line plots the maximum of the orbital period and Eq. (\ref{eq:tdis}) with $M_{\rm WD} = 0.6M_{\odot}$. Top panels show discs of mass $M_{\rm disc} = 10^{23} \ \rm g$, while bottom panels have more massive discs with $M_{\rm disc} = 10^{23} \ \rm g$. Left panels show an orbital pericentre of $q=R_{\odot}$, while the right panels show a larger pericentre of $q = 15 R_{\odot}$. In each panel the disc profile exponent is $\beta = 1.5$ for Kuiper Belt object orbits (solid blue line) and Oort cloud orbits (dot-dashed red line), and $\beta=2$ for Kuiper Belt object orbits (dashed green line) and Oort cloud orbits (dotted cyan line). 
    }
    \label{fig:times1}
\end{figure*}

Consider now a body of size $R_p$, and density $\rho_p$, but on a bound orbit with semi-major axis $a$ and eccentricity $e$. The pericentre distance is $q=a(1-e)$. During the passage through the gaseous disc, most of the energy loss occurs near pericentre. Hence, the energy loss is \citep{grietal2018}
\begin{equation}
\Delta E=-\frac{\pi}{2}C_D R_p^2 \Sigma_g(q) v_{\rm rel}^2 
.
\end{equation}
For a highly eccentric orbit, the velocity is approximated by the escape velocity $v_{\rm rel}^2 = 2GM_{\rm WD}/q$, and thus the energy loss is 
\begin{equation}
\Delta E=-\pi C_D GM_{\rm WD}R_p^2 \frac{\Sigma_g(q)}{q}  
.
\end{equation}
Since the orbital energy $E_{\rm orb} = -GM_{\rm WD}m_p/2a$ is small, the ratio between the dissipated energy and orbital energy, namely
\begin{equation}
\left|\frac{\Delta E}{E_{{\rm orb}}}\right|=\frac{3C_{D}\Sigma_{g}(q)}{2\rho_{p}R_{p}}\frac{a}{q}
,
\end{equation}
is small for very eccentric orbits. The body will be circularized and embedded in the gaseous disc within an orbital timescale if $R_p \lesssim (a/q) \Sigma_g/\rho_p$. This relation is definitely true for all Oort-cloud objects with extreme ratios of $a/q\gtrsim 10^8$, and is true for exo-Kuiper Belt objects under $\lesssim 1\ \rm km$ in size, where $a/q \approx 10^4$. 

For massive planetesimals and less massive gaseous disc (but much more massive than the planetesimals' mass, $M_{\rm disc} \gg \rho_p R_p^3$), the orbital energy will be dissipated only after several orbital cycles. The energy dissipation time can be defined as $\tau_E \equiv E_{\rm orb}/(dE/dt)$, where $dE/dt = \Delta E/P_{\rm orb}$, with $P_{\rm orb}$ being the orbital time of the body around the WD. The dissipation time is then
\begin{align} \label{eq:tdis}
\tau_{E} & = \frac{E_{{\rm orb}}}{dE/dt}=\frac{GM_{{\rm WD}}m_{p}}{2a}\frac{2\pi}{\Delta E}\sqrt{\frac{a^{3}}{GM_{{\rm WD}}}} \\ \nonumber
	& = \frac{q}{v_{K}(a)}\frac{4\pi\rho_{p}R_{p}}{3C_{D}\Sigma_{g}(q)}, 
\end{align}
where $v_{K}(a)\equiv(GM_{{\rm WD}}/a)^{1/2}$ is the Keplerian velocity of a circular orbit at semimajor axis $a$.

Figure \ref{fig:times1} shows the dissipation times for exo-Kuiper Belt object orbits and exo-Oort type orbits. The initial semimajor axes are $40\ \rm au$ for exo-Kuiper Belt object orbits and $10^4 \ \rm au$ for exo-Oort cloud orbits. The pericentre is either $q=R_{\odot}$ (left panels) or $q=15 R_{\odot}$ (right panels). In order to compute the timescale, for each value of $R_p$ we plot the maximum of the orbital time and $\tau_E$ from Eq. (\ref{eq:tdis}). We display the results for relatively low mass gaseous discs $M_{\rm disc} = 10^{23}\ \rm g$ (top panels) and a gaseous disc of mass $M_{\rm disc} = 10^{25}\ \rm g$ (bottom panels). Even for the lower mass disc, the dissipation times extend beyond the orbital periods only for large objects.

Therefore, the flat regions of the curves in Fig. \ref{fig:times1} reveal that most Oort-cloud planetesimals are captured within one orbital period. For extremely low-mass discs and orbital pericenters of $15 R_{\odot}$, planetesimals smaller than $0.1 (10)$ km are captured in one orbit in a $10^{23} (10^{25})$ g disc. Although larger objects are captured after longer times, the dissipation timescales are typically short relative to the orbital period. Thus, the capture rate mostly depends on the supply rate of Oort-cloud planetesimals into low angular momentum orbits, which is the main focus of the next section.

\section{Loss cone dynamics for exo-Oort clouds} \label{lc}

Due to the short dissipation timescales, planetesimals on low angular momentum (high-$e$) orbits are  captured within one orbtial time and effectively ``lost''. Therefore the capture rate is dominated by the flux of planetesimals into high-$e$ orbits,  which is generally dominated by collective diffusive processes. The high-$e$ (low angular momentum) orbits which are captured form the shape of a cone in action space. Therefore, the dynamical origins of the supply rate into high-$e$ orbits is denoted `loss cone dynamics' (see \citealp{mer13} for a review).

Loss cone dynamics are determined by two parameters (or times): The critical angular momentum required for loss, $J_c$ and the average change in the angular momentum within one orbit, $\Delta J$. If the change is small, then $|\Delta J| \ll J_c$, the replenishment time is longer than the orbital period, and the cone is mostly empty. This case corresponds to the empty loss cone regime and the loss rate is determined by the replenishment rate. Conversely, if  the change in the angular momentum per orbit is large, then $|\Delta J| \gtrsim J_c$, the replenishment time is much faster, and the loss rate is determined by the removal time. Here, the cone is mostly full: this situation corresponds to the full loss cone regime.

The drivers for the diffusive flux rate of exo-asteroids and exo-Oort cloud planetesimals are very different. Exo-asteroid diffusion rates are highly sensitive to the architecture of a planetary system and the details of its destabilization. Given the variations between the number of, mass of and mutual distances between exo-planets in separate systems, obtaining an analytical description of the loss cone regime for exo-asteroids across all systems is not yet feasible.

However, the situation is more optimisitc for exo-Oort cloud comets because their  diffusive flux rate is dominated by Galactic tides \citep{ht86}. Encounters of molecular clouds enhance the Oort cloud evolution and accretion rate \citep{wicnap2008}. Similar processes occur in the Galactic centre, with various types of relaxation, as well as massive perturbers and spiral-arms \citep{peretal2007,hamper2017}. 

In order to estimate the rate of captured exo-Oort cloud objects, we use the loss cone flux derived by \cite{ht86}.  The time to capture the planetesimals is bounded from above by the orbital time ($\sim 10^6 \rm yr$), while the replenishment time is on the order of Gyr. Therefore, we use the expression for the empty loss cone regime.

For a comet at semimajor axis $a$ and eccentricity $e$, the canonical actions are $L=\sqrt{\mu a}$, $J=\sqrt{\mu a(1-e^{2})}$ with $\mu\equiv GM_{{\rm WD}}$. For large enough semimajor axis $a$, Galactic tides are fast enough to perturb the orbits into very large eccentricity  (but still much slower than the orbital period, hence the empty loss cone assumption in maintained) and pericentre $q = a(1-e) \ll a $, which results in changing the action to $J_q \approx \sqrt{2 \mu q}$. For a critical pericenter $q_c$, orbits with $J_q \le J_c=\sqrt{2 \mu q_c}$ are destroyed. If $f_{\rm DF}(L,J)$ is the distribution function; namely that the number of Oort cloud comets in the action phase ${\rm d}L{\rm d}J$ is $dN = f_{\rm DF}(L,J){\rm d}L{\rm d}J$, \cite{ht86} find that the flux per unit $L$ is:

\begin{equation}
F_{\rm empty}(L) =f_{\rm DF}(L,J) \frac{160\pi^{3}G\rho_{0}}{3\mu^{2}}J_{q}L^4, \label{eq:lc}
\end{equation}

\noindent{}where $\rho_0 = 0.185 M_{\odot} \rm pc^{-3}$ is the local density. The transition to the empty loss cone occurs at $a\lesssim 10^4 \rm au$, a region which we assume to be devoid of planetesimals. Thus in our case the loss cone is always empty.

\subsection{Distribution function and rates}
Here we obtain an explicit distribution function and use it to calculate the loss rates of Oort cloud comets. 

We assume for simplicity that the distributions of the semimajor axis and the eccentricity are uniform, independent of each other, and are represented by power-laws as:
\begin{equation}
f_{A}(a)=k_{a}a^{-p};\ f_{E}(e)=1 \label{eq:distae}
\end{equation}
 where the domains are $a\in[a_{{\rm min}},a_{{\rm max}}];\ e\in[0,1)$.
The normalization is related to the distribution by $1=\intop_{a_{{\rm min}}}^{a_{{\rm max}}}f(a)da=1$.
For $p\ne1$ we have 
\begin{equation}
k_{a}=\frac{1-p}{a_{{\rm max}}^{1-p}-a_{{\rm min}}^{1-p}} \label{eq:norm}
\end{equation}
 while $k_{a}=1/\ln(a_{{\rm max}}/a_{{\rm min}})$ for $p=1$. 

If the joint distribution of $a$ and $e$ is independent, i.e. $f_{{\rm joint}}(a,e)=f_{A}(a)f_{E}(e),$
then the number of comets is 
\begin{equation}
dN=N_{{\rm tot}}f_{A}(a)f_{E}(e)=f_{{\rm DF}}(L,J)dLdJ. \label{eq:dn}
\end{equation}
Thus, the distribution function is given by the Jacobian transformation
(the off-diagonal terms vanish since $\frac{\partial L}{\partial e}=\frac{\partial a}{\partial J}=0)$
\[
f_{{\rm DF}}(L,J)=N_{{\rm tot}}f_{A}(a)f_{E}(e)\left|\frac{\partial a}{\partial L}\frac{\partial e}{\partial J}\right| \label{eq:transform}
.
\]
 By inverting the relations, we obtain $a=L^{2}/\mu$ and $e=1-J^{2}/L^{2}$.
The derivatives are 
\[
\frac{\partial a}{\partial L}=\frac{2L}{\mu};\ \frac{\partial e}{\partial J}=-2\frac{J}{L^{2}}
\]
yielding
\begin{align} \label{eq:df}
f_{{\rm DF}}(L,J) & =4\frac{N_{{\rm tot}}}{\mu}k_{a}a(L)^{-p}\frac{J}{L}\\ \nonumber
 & =4\mu^{p-1}N_{{\rm tot}}k_{a}\frac{J}{L^{2p+1}}. 
\end{align}
It can be verified that the number of comets is retained, namely $N_{\rm tot} = \intop_{L_{{\rm min}}}^{L_{{\rm max}}}\intop_{0}^{J_{{\rm max}}(L)}f_{\rm DF}(L,J)dJdL$.

For the empty loss case, by using Eqns (\ref{eq:lc}) and (\ref{eq:df}), we find that the empty loss cone rate is
\begin{equation}
F(L)=\frac{640\pi^{3}G\rho_{0}}{3}\mu^{p-3}N_{{\rm tot}}k_{a}J_q^{2}L^{3-2p}
.
\end{equation}
Note that $F(L)$ is independent of $L$ if $p=3/2.$ After plugging
in, e.g. $p=3/2$, $a_{{\rm in}}=10^{4}$au and $a_{{\rm out}}=10^{5}$au,
$N_{{\rm tot}}=10^{9},$ $q=R_{\odot}$, we obtain
\begin{equation}
F(L)  = 7.8\cdot10^{-3}\left(\frac{q}{R_{\odot}}\right)\left(\frac{N_{{\rm tot}}}{10^{9}}\right)\left(\frac{a_{{\rm max}}}{10^{5}{\rm AU}}\right)^{1/2}{\rm yr}^{-1}
.\label{eq:nice_r}
\end{equation}

\subsection{Size-dependent rate}

In deriving Eq. (\ref{eq:nice_r}) we assumed that all comets have a single mass. In general, however, the comets are described by a mass function $dN/dm\propto m^{-l}$. The overall
number of comets with mass $m$ is \citep{grietal2018}
\[
N(m)=\frac{2-l}{l-1}\frac{M_{{\rm tot}}}{m^{l-1}m_{{\rm up}}^{2-l}}
\]
where $1>l>2$. 

For a collisional distribution \citep{d69} , $l=11/6$, total mass of $M_{{\rm tot}}=10^{28}{\rm g}$,
and the largest body has a mass of $m_{\rm up}\approx10^{21}$g, which corresponds to a $\sim100$
km planetesimal, We have 
\[
N(m)=2\cdot10^{7}\left(\frac{m}{10^{21}{\rm \ g}}\right)^{-5/6}
.
\]
For the size dependence, we use $m=4\pi\rho_{p}R_{p}^{3}/3$ to obtain,
for $\rho_{p}~=~1$~${\rm g\ cm^{-3}}$,
\[
N(R_{p})=6\cdot10^{6}\left(\frac{R_{p}}{100\ {\rm km}}\right)^{-5/2}
.
\]
Finally, for the empty loss cone regime, we obtain the size-dependent analog to equation (\ref{eq:nice_r}) 
\begin{equation}
F(L,R_{p})=1.4\cdot10^{-2}\left(\frac{R_{p}}{10\ {\rm km}}\right)^{-5/2}\left(\frac{q}{R_{\odot}}\right)\left(\frac{a_{{\rm max}}}{10^{5}{\rm AU}}\right)^{1/2}{\rm yr}^{-1}
.
\label{FLRp}
\end{equation}
The total captured mass rate is 

\begin{equation}
\frac{dm}{dt} = \int_{R_{p} = 0 \ {\rm km}}^{100 \ {\rm km}} 4\pi R_p^2 \rho_{p} F(R_p) dR_p \approx 10^{18} \ {\rm g \ yr^{-1}}.
\end{equation}

 For the assumed Oort-cloud mass of $M_{\rm tot} = 10^{28}\rm g$, up to $10$ per cent of the mass could be captured within 1 Gyr for $q=R_{\odot}$, whereas numerical simulations show that $\sim 10^{-5}$ of the mass is disrupted, the analytical calculation of \cite{alcetal1986} finds that $\sim 1 \% $ of the objects are disrupted.

The latter is only an averaged estimate and depends on many parameters. The number density of the cloud reduces over time. In addition, the Galactic tide works over $\sim$ Gyr timescales, while comets have taken their new orbits only after stellar mass loss and the birth of the WD. Moreover, the furthest away planetesimals will be depleted first. Therefore, the maximal excursion will also change over time.

Regardless of the uncertainties, the calculated rates are comparable and provide an analytic framework to estimate the order of magnitudes. In addition, the loss rate depends linearly on the critical pericenter, $q_{\rm c}$. Thus, extended gaseous discs far beyond the Roche limit, $r_{\rm out} \gg R_{\odot}$, will capture planetesimals by a factor $r_{\rm out}/R_{\odot}$ higher than the number of planetesimals which are disrupted, regardless of their sizes and disc properties (up to some limit). In our case of an outer WD radius up to $20 R_{\odot}$, the embedded mass could be $\sim 20$ times the mass of the disrupted planetesimals, significantly altering the gas-to-dust ratio of the disc.









\section{Discussion} \label{sec5}

Our results potentially have implications for all WD planetary debris discs, not just the ones with embedded planetesimals. The pebbles and boulders captured change the size distribution of the debris disc, with implications for both modelling and observations. From the modelling perspective, a time-dependent size distribution and steady, periodic or intermittent inflow of material should be incorporated into simulations; a collisional cascade code like that used by \cite{kenbro2017a, kenbro2017b} can accomplish this task. 

From the observational perspective, there is increasing evidence of WD debris discs showcasing variability on the per cent to tens of per cent level on yearly timescales
\citep{ganetal2008,wiletal2014,xujur2014,faretal2018,xuetal2018,swaetal2019}. Not all these discs have observable gas, although it might be present anyway \citep{bonetal2017}. Flux changes over yearly or decadal timescales may be the result of changing size distributions and orbital shifting within those discs. Attempting to reproduce these changes in a particular system due to small-body capture represent enticing challenges for dedicated future studies.

One particularly interesting result of our investigations is that for sufficiently extended  gaseous discs, capture is more likely than disruption (see end of Section \ref{lc}). Consequently, there are implications for how debris discs are recycled on timescales much longer than the $\sim$ Myr upper bound suggested by \cite{giretal2012}. There are several important consequences. The first is a new source-sink relationship between the dust and the gas in the disc. Another are the geometric signatures resulting from a captured planetesimal \citep{manetal2019}. A third is the possibility of capture of multiple large planetesimals which could gravitationally perturb one another as well as the disc. A fourth is captured objects which could reside close to the locations where second-generation planetesimals may form from a massive disc \citep{vanetal2018}.

We assumed a fixed gaseous disc mass (except in Section 3.2).  The ISM planetesimals have a large (positive) energy, therefore their capture rate depends both  on their orbits and on the gasous disc properties. The capture rate of ISM planetesimals has a shallow dependence on the gaseous disc mass and surface density profile (see Eq. \ref{eq:kkk}).  Loosely bound exo-KBO and exo-Oort orbits have low orbital energy, and generally their energy dissipation time is shorter than their long orbital time (see Fig. \ref{fig:times1}), therefore for most cases their capture rate is independent of the disc mass and determined by the replenishment rate  of of new exo-KBO and exo-Oort comets into loss cone orbits, up to the point where there is not enough mass to capture the planetesimal within one passage (see Fig. \ref{fig:times1}).

Our key assumption is that the gaseous disc is extended beyond the Roche limit, and that there is a co-existent state of gas and solids (see Section \ref{s2}). The co-existent state is supported both by theory and observations, while the large radial extent is supported by theory, but currently not observed. Future observations should confirm or falsify our assumptions, and a deeper understanding of the physical mechanisms that permit the co-existent state is still lacking. By truncating the disc extent to the Roche limit, the planetesimals will be disrupted rather than captured. Capture would be possible only for compact planetesimals with high internal strength, which could provide the origin of the planetary core orbiting SDSS J1288+1040 \citep{manetal2019} well within the Roche limit.

 There is increasing evidence that just about all metal pollution arises from disc accretion, regardless if the discs can be seen \citep{beretal2014,farihi2016,bonetal2017}. The gaseous components of the disc can enhance the supply rate of solids that eventually form the debris component of the disc.  The link between the chemical composition of the embedded seed and the eventual abundances measured in the white dwarf atmosphere is tenuous due to seed's acquisition of solids, chemical mixing with the existing disc, and subsequent fragmentation before accretion onto the white dwarf.

 The chemical composition of the existing discs themselves \citep{reaetal2005,reaetal2009} are produced from the destruction of minor planets. Although these minor planets are predominantly dry \citep{juryou2014,holetal2018}, in contrast \cite{xuetal2017} discovered a volatile-rich metal polluted WD with a nitrogen mass fraction comparable to that of comet Halley and higher than than that of any inner solar system object. This metal polluted WD also contains significant fractions of other volatiles like carbon which are more characteristic of cometary objects than terrestrial rocky objects. Combined with the water-rich progenitor minor planets in a few other WD systems \citep{faretal2013,radetal2015,genetal2017}, the volatile-rich object from \cite{xuetal2017} suggests that multiple reservoirs supply WDs with metal pollutants. Water retention in minor planets is possible throughout the giant branch stages of stellar evolution \citep{malper2016,malper2017a,malper2017b}.

Finally, our numerical results are subject to the usual caveats of the currently unconstrained nature of architectures of WD planetary systems and the unknown disc lifetimes. However, we deliberately have kept our formalism general enough to allow the reader to insert their favoured distributions, rates and orbits into the equations. 

\section{Summary} \label{sec6}
The dynamic environment close to WDs is strongly influenced by external factors, with incoming exo-planetesimals representing frequent interlopers. After one of them disrupts at the WD Roche radius and forms a disc, the outward spread of the resulting gas allows new exo-planetesimals with orbital pericentres outside of the Roche radius to interact with the disc without being destroyed.

Sometimes this interaction leads to direct capture, representing the scenario we study here. Enabling this work is the analytical formalism developed by \cite{grietal2018}, which ultimately led to explicit expressions for capture rates of interstellar planetesimals (equation \ref{eq:rate}) and orbital dissipation timescales for the already-bound orbits of exo-Kuiper belts and exo-Oort cloud comets (equation \ref{eq:tdis}). Estimating the fraction of exo-Kuiper belt objects which are on capture orbits is currently too unconstrained to be analytically characterised. However, this fraction may be computed for exo-Oort cloud comets (equation \ref{FLRp}) by using the supply rate from Galactic tides \citep{ht86}. These equations are general enough to be applied to most white dwarf planetary systems with discs.

We found that capture of interstellar objects into white dwarf discs is negligible, but capture from (already bound) exo-Kuiper belt and exo-Oort cloud objects is important.  In an extended gaseous disc of $\sim 20R_{\odot}$ capture is $\sim 20$ times more likely than disruption. The probability of capture scales strongly with planetesimal size, and there is a nonzero probability that the planetary core fragment embedded with the SDSS J1288+1040 disc was captured by the gaseous component of that disc. Besides minor planets which are at least a km in size, capture of much smaller m- and cm-size particles like boulders and pebbles is likely to provide a continuously changing size distribution in the discs. Capture on yearly or decadal timescales might even explain variability which is seen in some WD discs.

\section*{Acknowledgements}
 We thank the referee for their useful feedback, which has improved the manuscript.
Both EG and DV acknowledge useful discussions at the Lorentz Workshop {\it Trendy-2} which has led to this result. We also thank Brian David Metzger and Roman Rafikov for helpful correspondence. EG acknowledges support by the Technion Irwin and
Joan Jacobs Excellence Fellowship for outstanding graduate students. DV gratefully acknowledges the support of the STFC via an Ernest Rutherford Fellowship (grant ST/P003850/1).






\bsp	
\label{lastpage}
\end{document}